# Extracting the Most Weighted Throughput in UAV Empowered Wireless Systems With Nonlinear Energy Harvester

Yanjie Dong, *Student Member, IEEE*[‡], Julian Cheng, *Senior Member, IEEE*[†],
Md. Jahangir Hossain, *Senior Member, IEEE*[†], and Victor C. M. Leung, *Fellow, IEEE*[‡]

[‡]Department of Electrical and Computer Engineering, The University of British Columbia, Vancouver, BC, Canada
[†]School of Engineering, The University of British Columbia, Kelowna, BC, Canada
Emails: {ydong16, vleung}@ece.ubc.ca, {julian.cheng, jahangir.hossain}@ubc.ca

*Abstract*—With the maturity of unmanned aerial vehicle (UAV) technology, this work investigates the integration of UAV into wireless communication systems. Since the UAV is powered by a capacity-limited battery, this work proposes to use the radio energy harvesting technology at the UAV in order to extend the lifetime of UAV empowered base station. To extract the most weighted throughput of UAV empowered wireless systems, the dirty paper coding scheme and information-theoretic uplink-downlink channel duality are exploited to propose an extracting the most weighted throughput algorithm. Numerical results are used to verify the proposed algorithm.

## I. Introduction

Unmanned aerial vehicles (UAVs) have received huge attention for various applications, such as hazardous region surveillance and regional imaging in the past several decades [1]–[3]. As an important use case of UAVs, the UAV empowered wireless communication becomes an emerging topic recently due to the advantages of fast and flexible deployment as well as high probability of line-of-sight (LoS) channels [2]–[5]. For example, when the terrestrial base station fails, a UAV empowered base station (UAV-BST) can be launched to serve the users in outage. Moreover, the UAV-BST can also offload the crowded terrestrial base station such that more users can be accessed in the network simultaneously [1].

On the roadmap of intergrating UAV technology into wireless communication systems, several research progress has been reported [4]–[9]. The authors in [4] investigated the application of UAV enabled backhaul connections for the small cell base stations. Moreover, they suggested to use proactive caching strategy in order to enhance the quality of experience of users [4]. The authors in [5] studied the design of trajectory, transmit power and user association for the throughput maximization for UAV-enabled multi-user downlink communication. The authors in [6] investigated the energy consumption minimization problem while guaranteeing the communication quality of service in a UAV based mobile cloud computing system, where UAV-BST is able to support the requirement of users offloaded by the terrestrial base station. The authors in [7] investigated the three-dimensional placement of a single UAV-BST to absorb the users from the terrestrial base station based on the probabilistic LoS channel model. Then, the authors in [8] developed a particle-swarm based algorithm to minimize the number of active UAV-BSTs in predetermined locations.

While the aforementioned literatures mainly focus on the design of communication protocol, trajectory and placement of UAVs, the wireless powered communications, which can extend the lifetime of UAV after per charge, have not been well investigated in the UAV empowered wireless communication systems. In this work, we focus on extracting the most throughput via dirty paper coding scheme (DPCS) in the UAV empowered wireless communication systems with practical nonlinear energy harvester. Since a UAV is usually powered by a capacity-limited battery, the cruising endurance of the UAV is restricted. In order to extend the endurance of UAV, we propose to equip the UAV with a practical nonlinear energy harvester such that the UAV reduces the usage of battery energy during information transmission. Our major contributions are two folds:

- We investigate the integration of UAV into wireless communication systems with a practical nonlinear energy harvester.
- We propose an optimal algorithm to maximize the weighted throughput of the UAV empowered wireless communication systems.

The remaining work is organized as follows. Section II describes the system model and formulates the optimization problem of the UAV empowered wireless communication systems with a nonlinear energy harvester. In Section III, we propose an extracting the most weighted throughput (EMWT) algorithm for the formulated problem. Numerical results are presented in Section IV, and conclusions are drawn from Section V.

*Notations:* Vectors and matrices are shown in bold lower-case letters and bold uppercase letters, respectively. $\mathbb{C}^{N \times M}$ and $\mathbb{H}^N$ respectively denote all the $N \times M$ dimension complex value matrices and $N \times N$ Hermitian matrices. $\|\cdot\|_{\text{F}}$ and $|\cdot|$ are the Frobenius norm and absolute value of a scalar (or determinant of a matrix), respectively. $\sim$ stands for "distributed as". $\boldsymbol{I}_N$ and $\boldsymbol{1}_{N \times M}$ denote, respectively, an $N$ dimensional identity matrix and an all zero matrix with $N$ rows and $M$ columns. The expectation of a random variable is denoted as $\mathbb{E}[\cdot]$. vec$[\boldsymbol{W}]$ obtains a vector by stacking the columns of $\boldsymbol{W}$ under the other. $\{\boldsymbol{w}_n\}_{n \in \mathcal{N}}$ represents the set made of $\boldsymbol{w}_n$, $n \in \mathcal{N}$. For a square matrix $\boldsymbol{W}$, $\boldsymbol{W}^{\text{H}}$ and Tr$(\boldsymbol{W})$ denote its conjugate transpose and trace, respectively. $\boldsymbol{W} \succeq \boldsymbol{0}$ and $\boldsymbol{W} \succ \boldsymbol{0}$ respectively denote that $\boldsymbol{W}$ is a positive semidefinite and $\boldsymbol{W}$ is a positive definite matrix.

## II. SYSTEM MODEL AND PROBLEM FORMULATION

### A. Overall Description

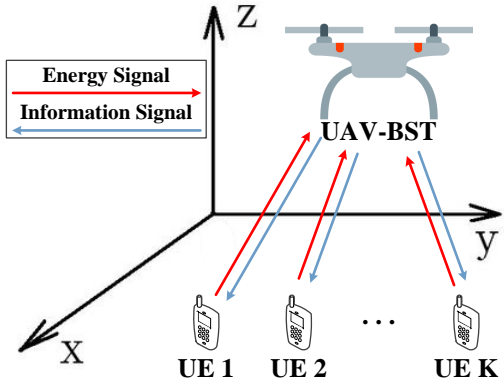

Fig. 1. An illustration of the UAV empowered wireless system with nonlinear energy harvester, where the UAV is equipped with a nonlinear energy harvester.

We consider a UAV empowered wireless system which consists of a UAV-BST and a set of user equipments (UEs) as shown in Fig. 1. Let $\mathcal{K} = \{1, 2, \ldots, K\}$ denote the set of UEs. The $k$-th UE is equipped with $N$ transmit antennas, and the UAV-BST is equipped with one single antenna. We propose a protocol that all the UEs transmit energy signals to UAV-BST and receive information from UAV-BST in the uplink and downlink period, respectively. Here, the energy consumed by UAV-BST in downlink transmission is harvested during the uplink period. Since the battery of UAV has limited capacity, the proposed protocol can extend the cruising endurance of the UAV. We consider that the UAV-BST communicates with UEs via a frame-based character with unit duration for each frame. Hence, the terminologies "energy" and "power" can be used interchangeably. We assume that the UAV empowered operates in time-division-duplex mode; therefore, the UAV-BST and UEs have the perfect channel state information via exploiting the uplink-downlink reciprocity.

Let $\boldsymbol{h}_{k,\text{UL}}$, $\boldsymbol{w}_k$ and $x_{k,\text{UL}}$ respectively denote the vector of complex channel coefficients, transmit beamforming vector and transmit energy symbol of the $k$-th UE in the uplink period, the received signal at UAV-BST is denoted as

$$y_{\text{UAV}} = \sum_{k=1}^{K} \boldsymbol{h}_{k,\text{UL}}^{\text{H}} \boldsymbol{w}_k x_{k,\text{UL}} + z_{\text{UAV}} \quad (1)$$

where $z_{\text{UAV}} \sim \mathcal{CN}(0, \sigma^2)$ is the additive white Gaussian noise (AWGN) at UAV-BST. Here, the transmit symbol $x_{k,\text{UL}}$ follows a circularly symmetric complex Gaussian distribution as $x_{k,\text{UL}} \sim \mathcal{CN}(0,1)$.

Ignoring the power in the AWGN, the amount output power at the nonlinear energy harvester is denoted as

$$P_{\text{UAV}}^{\text{OUT}} = \frac{c}{1 - M_{\text{UAV}}} \left( \frac{1}{1 + \exp(-a(P_{\text{UAV}}^{\text{IN}} - b))} - M_{\text{UAV}} \right) \quad (2)$$

with

$$P_{\text{UAV}}^{\text{IN}} = \sum_{k=1}^{K} \left| \boldsymbol{h}_{k,\text{UL}}^{\text{H}} \boldsymbol{w}_k \right|^2 \quad (3)$$

and

$$M_{\text{UAV}} = \frac{1}{1 + \exp(ab)} \quad (4)$$

where the constants $a$, $b$ and $c$ are the shaping parameters of the nonlinear energy harvester [10], [11].

Let $\boldsymbol{h}_{k,\text{DL}}$ denote the vector of complex channel coefficients for the $k$-th UE in the downlink period, the received signal at the $k$-th UE is denoted as

$$\boldsymbol{y}_{k,\text{DL}} = \boldsymbol{h}_{k,\text{DL}} \sum_{n=1}^{K} \sqrt{p_n} x_{n,\text{DL}} + \boldsymbol{z}_{k,\text{DL}} \quad (5)$$

where $p_n$ and $x_{n,\text{DL}}$ respectively denote the transmit power and transmit symbol for the $n$-th UE; and $\boldsymbol{z}_{k,\text{DL}} \sim \mathcal{CN}(\boldsymbol{0}, \sigma^2 \boldsymbol{I}_N)$ denotes the AWGN at the $k$-th UE. Here, the transmit symbol $x_{n,\text{DL}}$ follows a circularly symmetric complex Gaussian distribution as $x_{n,\text{DL}} \sim \mathcal{CN}(0,1)$.

Since the DPCS can achieve the maximum throughput via sequentially encoding each UE without interference from the previous encoded UEs [12], [13], the DPCS enabled weighted throughput is obtained as

$$R_{\text{DL}} = \max_{\boldsymbol{\pi}} \sum_{k=1}^{K} w_k \log \left( \frac{\left| \boldsymbol{I}_N + \sum_{n=1}^{k} \frac{p_{\pi_n}}{\sigma^2} \boldsymbol{H}_{\pi_n,\text{DL}} \right|}{\left| \boldsymbol{I}_N + \sum_{n=1}^{k-1} \frac{p_{\pi_n}}{\sigma^2} \boldsymbol{H}_{\pi_n,\text{DL}} \right|} \right) \quad (6)$$

where $\boldsymbol{\pi} = \{\pi_1, \pi_2, \ldots, \pi_K\}$ denotes a encoding permutation with $\pi_n$ as the sequence of the $n$-th UE. Here, the constant $w_k$ is the weight of throughput for the $k$-th UE, and $\boldsymbol{H}_{\pi_n,\text{DL}} \triangleq \boldsymbol{h}_{\pi_n,\text{DL}} \boldsymbol{h}_{\pi_n,\text{DL}}^{\text{H}}$. In addition, the utilized power of UAV-BST is confined by the harvested power during the uplink phase as

$$\frac{1}{\phi} \sum_{k=1}^{K} p_{\pi_k} + P_{\text{CIR}} \leq P_{\text{UAV}}^{\text{OUT}} \quad (7)$$

where $P_{\text{CIR}}$ is the circuit power consumption of UAV-BST during the downlink transmission.

### B. Problem Formulation

Our objective is to maximize the DPCS enabled weighted throughput via joint design of uplink beamforming and downlink power control. Therefore, the DPCS enabled weighted

throughput maximization problem is formulated as

$$\max_{\substack{\{\boldsymbol{w}_{\pi_k},p_{\pi_k}\}_{k\in\mathcal{K}} \\ \boldsymbol{\pi}}} \sum_{k=1}^{K} w_k \log\left(\frac{\left|\boldsymbol{I}_N + \sum_{n=1}^{k} \frac{p_{\pi_n}}{\sigma^2}\boldsymbol{H}_{\pi_n,\text{DL}}\right|}{\left|\boldsymbol{I}_N + \sum_{n=1}^{k-1} \frac{p_{\pi_n}}{\sigma^2}\boldsymbol{H}_{\pi_n,\text{DL}}\right|}\right) \quad (8\text{a})$$

$$\text{s.t.} \quad \frac{1}{\phi}\sum_{k=1}^{K} p_{\pi_k} + P_{\text{CIR}} \leq P_{\text{UAV}}^{\text{OUT}} \quad (8\text{b})$$

$$\|\boldsymbol{w}_{\pi_k}\|_{\text{F}}^2 \leq P_{\pi_k}^{\max}, \forall k \quad (8\text{c})$$

$$p_{\pi_k} \geq 0, \forall k \quad (8\text{d})$$

where the constraint in (8b) is the overall power consumption constraint of UAV-BST in the downlink period, and the constant $P_{\pi_k}^{\max}$ of constraints in (8c) denotes the maximum transmit power of the $\pi_k$-th UE.

***Remark 1:*** We observe that the optimization problem (8) is challenging to solve due to the permutation variables $\boldsymbol{\pi}$ and non-convex objective function (8a). In order to deal with these two issues, we leverage the information-theoretic uplink-downlink duality to reduce the permutation variables in $\boldsymbol{\pi}$ and to transform the non-convex objective function (8a) into a convex one.

## III. THROUGHPUT EXTRACTION VIA JOINT DESIGN OF UPLINK BEAMFORMING AND DOWNLINK POWER CONTROL

We select an encoding permutation $\boldsymbol{\pi}^*$ such that $w_{\pi_1^*} \geq w_{\pi_2^*} \geq \ldots \geq w_{\pi_K^*}$. It is shown in [14] that the permutation $\boldsymbol{\pi}^*$ is optimal. With the selected permutation $\boldsymbol{\pi}^*$, we exploit the information-theoretic uplink-downlink duality where the dual uplink channel of (5) is obtained as [12]

$$\widetilde{\boldsymbol{y}}_{\text{UL}} = \sum_{k=1}^{K} \widetilde{\boldsymbol{h}}_{\pi_k^*,\text{UL}} \sqrt{p_{\pi_k^*}} x_{\pi_k^*,\text{DL}} + \boldsymbol{z}_{\text{UL}} \quad (9)$$

where $\boldsymbol{z}_{\text{UL}} \sim \mathcal{CN}(\boldsymbol{0},\sigma^2 \boldsymbol{I}_N)$ denotes the dual AWGN, and $\widetilde{\boldsymbol{h}}_{\pi_k^*,\text{UL}} \triangleq \boldsymbol{h}_{\pi_k^*,\text{DL}}$.

Following [14, Lemma 1], the DPCS enabled weighted throughput (6) equals to the weighted throughput of the uplink channel in (9), which is shown as (10) with $\pi_{K+1}^* = 0$.

With the equivalent expression of DPCS enabled weighted throughput in (10), we not only remove the permutation from the optimization variable but also obtain a convex objective function. Therefore, the DPCS enabled weighted throughput maximization problem is equivalently transformed into (13).

In order to develop the optimal beamforming vector in the uplink period, we are motivated to exploit of the structure of problem (13). We observe that a larger harvested power of UAV-BST results in a larger weighted throughput in (13a). In addition, the output power of the nonlinear energy harvester (2) monotonically increases with the input power (3). We are motivated to perform the maximal ratio transmission in the uplink period by setting

$$\boldsymbol{w}_{\pi_k^*} = \sqrt{P_{\pi_k^*}^{\max}} \frac{\boldsymbol{h}_{k,\text{UL}}}{\|\boldsymbol{h}_{k,\text{UL}}\|_{\text{F}}}. \quad (11)$$

As a result, the maximum amount of harvested power at UAV-BST is obtained as

$$P_{\text{UAV}}^{\text{OUT}} = \frac{\frac{c}{1+\exp\left(-a\left(\sum_{k=1}^{K} P_{\pi_k^*}^{\max}\|\boldsymbol{h}_{k,\text{UL}}\|_{\text{F}}^2 - b\right)\right)} - cM_{\text{UAV}}}{1 - M_{\text{UAV}}}. \quad (12)$$

---

$$R_{\text{DL}} = R_{\text{UL}} \quad (10)$$

$$= \max_{\sum_{k=1}^{K}\frac{p_{\pi_k^*}}{\phi}+P_{\text{CIR}}\leq P_{\text{UAV}}^{\text{OUT}}} \sum_{k=1}^{K}\left(w_{\pi_k^*} - w_{\pi_{k+1}^*}\right)\log\left|\boldsymbol{I}_N + \sum_{n=1}^{k}\frac{p_{\pi_n^*}}{\sigma^2}\boldsymbol{H}_{\pi_n,\text{DL}}\right|$$

---

$$\max_{\{\boldsymbol{w}_{\pi_k^*},p_{\pi_k^*}\}_{k\in\mathcal{K}}} \sum_{k=1}^{K}\left(w_{\pi_k^*} - w_{\pi_{k+1}^*}\right)\log\left|\boldsymbol{I}_N + \sum_{n=1}^{k}\frac{p_{\pi_n^*}}{\sigma^2}\boldsymbol{H}_{\pi_n^*,\text{DL}}\right| \quad (13\text{a})$$

$$\text{s.t.} \quad \frac{1}{\phi}\sum_{k=1}^{K} p_{\pi_k^*} + P_{\text{CIR}} \leq P_{\text{UAV}}^{\text{OUT}} \quad (13\text{b})$$

$$\|\boldsymbol{w}_{\pi_k^*}\|_{\text{F}}^2 \leq P_{\pi_k^*}^{\max}, \forall k \quad (13\text{c})$$

$$p_{\pi_k^*} \geq 0, \forall k \quad (13\text{d})$$

---

$$\max_{\{p_{\pi_k^*}\}_{k\in\mathcal{K}}} \sum_{k=1}^{K}\left(w_{\pi_k^*} - w_{\pi_{k+1}^*}\right)\log\left|\boldsymbol{I}_N + \sum_{n=1}^{k}\frac{p_{\pi_n^*}}{\sigma^2}\boldsymbol{H}_{\pi_n^*,\text{DL}}\right| \quad (14\text{a})$$

$$\text{s.t.} \quad \frac{1}{\phi}\sum_{k=1}^{K} p_{\pi_k^*} + P_{\text{CIR}} \leq \frac{\frac{c}{1+\exp\left(-a\left(\sum_{k=1}^{K} P_{\pi_k^*}^{\max}\text{Tr}\left(\boldsymbol{H}_{\pi_k^*,\text{UL}}\right)-b\right)\right)} - cM_{\text{UAV}}}{1 - M_{\text{UAV}}} \quad (14\text{b})$$

$$p_{\pi_k^*} \geq 0, \forall k \quad (14\text{c})$$

Substituting the maximum amount of harvested power (12) into the right hand side of (13b) and dropping (13c), we obtain the simplified optimization problem as (14). We observe that the optimization problem (14) is convex with $\boldsymbol{H}_{\pi_k^*,\text{UL}} \triangleq \boldsymbol{h}_{\pi_k^*,\text{UL}} \boldsymbol{h}_{\pi_k^*,\text{UL}}^{\text{H}}$. Solving the problem (14) via off-the-shelf toolbox, e.g., CVX [15], we obtain the EMWT algorithm.

## IV. NUMERICAL RESULTS

In this section, we use numerical results to verify the system performance of the proposed EMWT algorithm. The simulation parameters are set as follows. Each UE is equipped with 3 antennas. Since channels contains LoS component, we use the Rician fading to model the mutlipath transmission between UEs and UAV-BST with the Rician factor is set as $\kappa = 2$. Therefore, the power of LoS and non-line-of-sight components satisfy $\frac{\kappa}{\kappa+1} d_k^{-\alpha}$ and $\frac{1}{\kappa+1} d_k^{-\alpha}$ where $d_k$ and $\alpha = 2.5$ are link distance of the $k$-th UE and pathloss exponent, respectively. The power of AWGN is set as 0.001 mW. For the nonlinear energy harvester, we set the parameters $a$, $b$ and $c$ as $a = 6400$, $b = 0.003$ and $c = 200$ mW [10]. The maximum transmit power $P_k^{\max}$ of each UE is set as 200 mW, and the amplifier efficiency $\phi$ of UAV-BST is set as 0.8. The vector of weights is set as $[0.3, 0.25, 0.2, 0.15, 0.1]$.

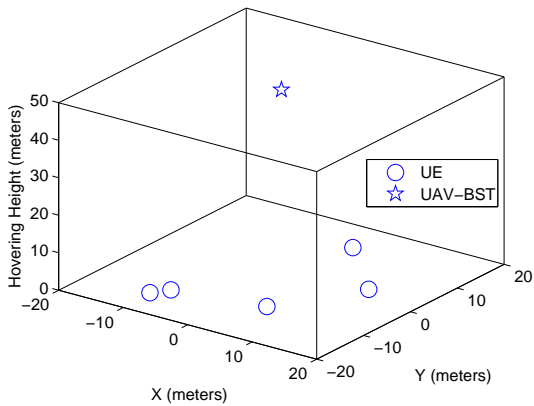

Fig. 2. The topology for the UAV empowered wireless communication system.

Figure 2 illustrates the topology for the UAV empowered wireless communication system. The UAV-BST hovers at a height of 50 meters and locates at the center of circle. The number of UEs is 5, and the distance between each UE and UAV-BST is uniformly distributed 10 to 20 meters from the center of circle.

Figure 3 shows the variation of system weighted throughput with the circuit power consumption of UAV-BST. We observe that the system weighted throughput decreases with the circuit power consumption of UAV-BST. For example, when the maximum output power of nonlinear energy harvester is 100 mW, the system weighted throughput can decrease at most 45.5% as the circuit power consumption increases from 40 mW to 80 mW. This is due to the fact a large circuit power consumption of UAV-BST results in a small amount of power can be used for information transmission in the downlink period. Thus, the reduced amount of available power consumption leads to a small system weighted throughput. From Fig. 3, we also observe that increasing the maximum output power of nonlinear energy harvester to 200 mW can compensate the system weighted throughput at the expense of hardware expenditure. Moreover, increasing the maximum output power of the nonlinear energy harvester can reduce the percentage of system weighted throughput degradation to 11.67%.

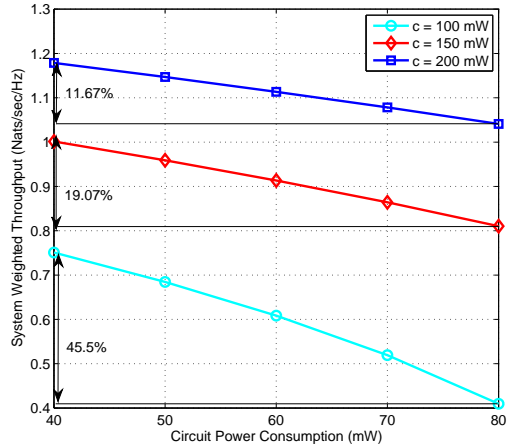

Fig. 3. The variation of system weighted throughput with the circuit power consumption of UAV-BST $P_{\text{CIR}}$.

## V. CONCLUSIONS

We investigated the system weighted throughput maximization problem in the UAV empowered wireless communication systems with nonlinear energy harvester. Via exploiting the DPCS and information-theoretic uplink-downlink channel duality, we proposed an EMWT algorithm to extract the most weighted throughput. Simulation results show that a larger circuit power consumption at the UAV-BST impairs the system weighted throughput. Moreover, the simulation results also demonstrate that increasing the maximum output power of nonlinear energy harvester can compensate the system weighted throughput at the expense of hardware expenditure.